\documentclass[final,5p,times,twocolumn]{elsarticle}
\usepackage{graphicx}
\usepackage{amsmath,amssymb,bm,amsfonts}
\usepackage{citesort}
\usepackage{slashed}
\newcommand{\Lagr}{\mathcal{L}}
\allowdisplaybreaks[1]
\begin{document}

\begin{frontmatter}
\title{{\boldmath$\eta$}-photoproduction in a gauge--invariant chiral unitary framework}

\author[Bonn]{Dino Rui\'c}
\author[Bonn]{Maxim Mai}
\author[Bonn,Julich]{Ulf-G. Mei{\ss}ner}

\address[Bonn]{Universit\"at Bonn,
             Helmholtz-Institut f\"ur Strahlen- und Kernphysik (Theorie)
         and Bethe Center for Theoretical Physics,
             D-53115 Bonn, Germany}

\address[Julich]{Forschungszentrum J\"ulich, 
             Institut f\"ur Kernphysik,
             Institute for Advanced Simulation, 
         and J\"ulich Center for Hadron Physics,
             D-52425 J\"ulich, Germany}
\date{\today}

\begin{abstract}
We analyse photoproduction of $\eta$ mesons off the proton in a
gauge-invariant chiral unitary framework. The interaction kernel 
for meson-baryon scattering is derived from the leading order chiral 
effective Lagrangian and iterated in a Bethe-Salpeter equation. 
The recent precise threshold data from the Crystal Ball at MAMI can be described 
rather well and the complex pole corresponding to the $S_{11}(1535)$ 
is extracted. An extension of  the kernel is also discussed.
\end{abstract}

\end{frontmatter}

\noindent{\textbf{Introduction}:}~
Coupled-channel unitary extensions of chiral perturbation theory
have been established as a viable tool to investigate the
chiral SU(3) dynamics of QCD, for early papers see e.g. 
\cite{Kaiser:1995eg,Oset:1997it,Nieves:1999bx,Oller:2000ma,Oller:2000fj}.
The main reason for this is that within such a framework, resonances close
to and even below the relevant thresholds can be generated dynamically. The
two premier examples are the $\Lambda (1405)$ that features prominently
in antikaon-proton scattering and the $S_{11}(1535)$ that dominates the
threshold cross section in eta photoproduction off protons. In the context
of unitarized chiral perturbation theory, this was first investigated
in Refs.~\cite{Kaiser:1995cy,Kaiser:1996js} based on the next-to-leading
order (NLO) effective Lagrangian (i.e. using the pertinent contact interactions).
The inclusion of p-waves was studied in \cite{CaroRamon:1999jf}, and the
role of the $\pi\pi N$ final-state was investigated in \cite{Inoue:2001ip}.
 $\eta$ and $\eta'$ photo- and electroproduction based on a U(3) extension
of the chiral Lagrangian was considered in \cite{Borasoy:2002mt}. In these
ground-breaking papers, the issue of gauge invariance was not considered.
In Ref.~\cite{Borasoy:2007ku}, a gauge-invariant framework
for meson photoproduction was developed and applied to associated strangeness production,
using the leading order Weinberg-Tomozawa interaction as the driving term in
the Bethe-Salpeter equation. In this note, we extend this method to the
process $\gamma p\to \eta p$ and in particular to a determination of the
mass and width of the  $S_{11}(1535)$. Eventually, NLO contributions will
have to be included, a first step for the purely hadronic interactions within
our approach was recently reported in \cite{Bruns:2010sv}.

\medskip

\noindent {\textbf{Framework}:}~
In this work we follow the construction procedure of the minimal 
approach to meson photoproduction, which is exactly unitary
in the subspace of meson-baryon channels
and 
gauge-invariant, as developed in Ref.~\cite{Borasoy:2007ku}. We refer 
the reader to that paper for a more detailed discussion. The first building 
block of the photoproduction amplitude is the meson--baryon 
interaction,  for which we consider the chiral effective Lagrangian of QCD at leading order:
	\begin{align}\label{eqn:LAGR}
	\Lagr^{(1)}_{\phi B}&=\langle \bar{B} (i\gamma_\mu D^\mu-m_0)B\rangle
	+\frac{D/F}{2}\langle \bar{B}\gamma_\mu \gamma_5[u^\mu,B]_\pm \rangle ~,
	\end{align}
where $\langle\ldots\rangle$ denotes the trace in flavor space, $D_\mu B 
=\partial_\mu B +[\Gamma_\mu,B]$,  $m_0$ is the baryon octet mass in the
chiral SU(3) limit, while $D$ and $F$ are the axial coupling constants. 
The relevant degrees of freedom are the Goldstone bosons described by the 
traceless  meson matrix $U=u^2=\exp(i{\phi}/{F_0})$, where $F_0$ is the 
meson decay constant in the chiral limit. The meson- and the low-lying 
baryon-fields are collected in traceless matrices $\phi$ and $B$,
respectively. Moreover, we use $u^\mu=i u^\dagger(\partial^\mu U
-i[v^\mu,U])u^\dagger$ the so-called chiral vielbein. The quark 
charge matrix is $Q={\rm diag}(2/3,-1/3,-1/3)$, and  the external vector field 
is encoded in $v^\mu=-eQA^\mu$. 
 
The expansion of the chiral connection $\Gamma^\mu=
[u^\dagger,\partial_\mu u]/2-i(u^\dagger v^\mu u+uv^\mu u^\dagger)/2$ 
in meson fields leads to the meson--baryon vertex of the leading chiral order, 
the Weinberg--Tomozawa (WT) term. However, at first chiral order, there are 
also the Born graphs, describing the $s$-channel and $u$-channel exchanges 
of an intermediate nucleon. The full inclusion of these graphs in the driving 
term of the Bethe-Salpeter equation (BSE) leads to conceptional and 
practical difficulties, which 
have not yet been solved to the best of our know\-ledge, 
see \cite{Bruns:2010sv,PCB:Diss,Mai:2011xx} for a more detailed discussion of this
issue. However, most chiral unitary approaches restrict their meson-baryon 
potential to this interaction, which generates the leading contribution 
to the s-wave scattering lengths. This approach has been remarkably 
successful in many cases,
thus we iterate the WT--potential $V_{WT}=g(\slashed q_1+\slashed q_2)$ via
the BSE in $d$ space-time dimensions to infinite order as follows
	\begin{align}\label{eqn:BSE}
	T(\slashed{q}_2, &\slashed{q}_1; p)= V_{WT}(\slashed{q}_2, \slashed{q}_1) 
	+\nonumber\\
	&\int\frac{d^d l}{(2\pi)^d}V_{WT}(\slashed{q}_2, \slashed{l}) 
	iS(\slashed{p}-\slashed{l})\Delta(l)T(\slashed{l}, \slashed{q}_1; p),
	\end{align}
where the in-, outgoing meson and the overall four-momentum are denoted by 
$q_1$, $q_2$ and $p$, respectively. The baryon and the meson propagators 
are represented by $iS(\slashed{p}) ={i}/({\slashed{p}-m+i\epsilon})$ 
and $i\Delta(k) ={i}/({k^2-M^2+i\epsilon})$, in order. Moreover, every element 
of the above equation is a matrix in the channel space. For the case of 
$\eta$-photoproduction considered here, we restrict ourselves to the following 
channels: $\{p \pi^0,~n \pi^{+},~p\eta,~\Lambda K^+,~\Sigma^0 K^+,~\Sigma^+ K^0\}$.

Because the kernel of the integral equation Eq.~(\ref{eqn:BSE}) stems from 
the contact interactions only, we have to deal at most with
one-meson-one-baryon loops, which are all dimensionally regularized 
throughout this work. It is not possible to express the terms necessary 
to absorb the divergencies in the BSE as counterterms derived from a local
Lagrangian. However, it is possible to alter the loop integrals in the
solution of the BSE in a way that is in principle equivalent to a proper 
modification of the chiral potential itself, see \cite{PCB:Diss}. In this 
spirit we apply the usual $\overline{MS}$ subtraction scheme, keeping in mind 
that the modified loop integrals are still scale-dependent. This
regularization scale ($\mu$) is used as a fitting parameter, reflecting the 
influence of higher order terms not included in our potential.

The functional form of the driving term allows to construct an explicit 
solution of the BSE, see \cite{Borasoy:2007ku}. Starting from the
corresponding scattering amplitude $T(\slashed{q}_2, \slashed{q}_1; p)$, 
where exact two-body unitarity is guaranteed by construction in the BSE framework, 
and following the recipe of \cite{Borasoy:2007ku} we are now able to construct 
the gauge-invariant photoproduction amplitude in a most natural way, without 
any use of ``artificial restoration'' of gauge invariance, i.e. adding
contact interactions, see e.g. \cite{Haberzettl:1998eq,Doring:2009qr}. In our approach, 
we simply couple the photon to any in- and external line as well as to 
the (momentum-dependent) vertices.

As an intermediate step we construct the amplitude $\Gamma$ for the process 
$p\to~B{}\phi$ starting from the potential derived from the leading order 
Lagrangian Eq.~(\ref{eqn:LAGR}) $\hat V=\slashed q \gamma_5 \hat g$, where 
$q$ is the outgoing meson momentum and $\hat g$ is the ($D-$, $F-$dependent) 
coupling constant corresponding to the second  term of Eq.~(\ref{eqn:LAGR}).  
Now we add the loop contribution that accounts for the final state interaction as follows
	\begin{align}\label{eqn:Gamma}
	 \Gamma(\slashed p,\slashed q) 
        = \hat V(\slashed q) + \int\frac{d^d l}{(2\pi)^d}T(\slashed{q},\slashed{l}; p) 
	iS(\slashed{p}-\slashed{l})\Delta(l)\hat V(\slashed{l}).\nonumber
	\end{align}
Note that the process in question is meson photoproduction off the proton, 
thus $\Gamma$, $\hat V$ and $\hat g$ are 6-vectors in the channel space.

\begin{figure}[h]
	\includegraphics[width=1.0\linewidth]{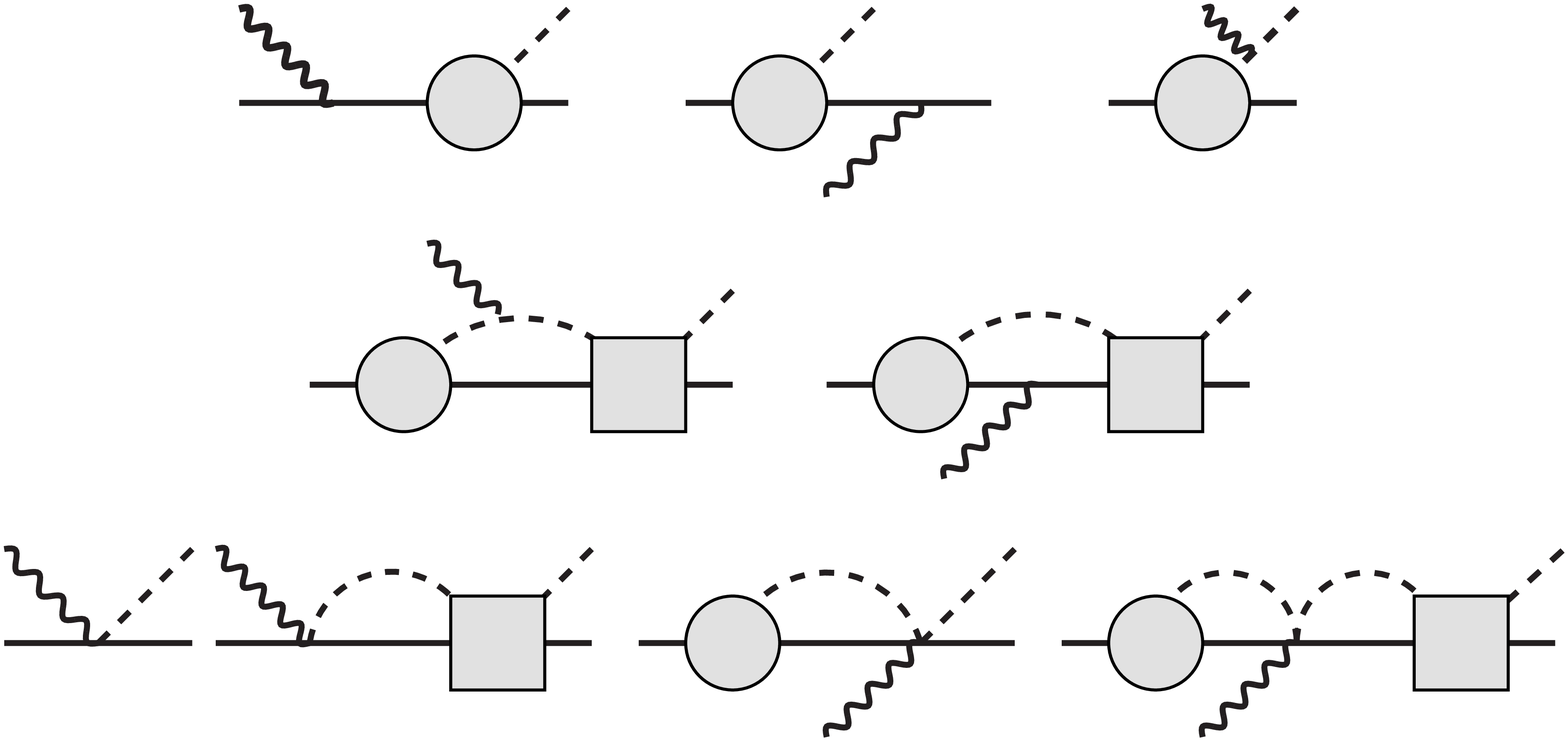}
        \caption{Different topologies contributing to gauge-invariant
        photomeson production off the proton. Solid, dashed and wiggly lines
        denote baryons, mesons and photons, in order. The filled
        circles/squares denote the meson production/the meson-baryon
        interaction, i.e. $\Gamma$ and $T$, respectively.
        \label{fig:topo}}
	\end{figure}

In the next step we couple the photon in every possible place to the hadronic 
skeleton. A photon can couple via: $B\phi\gamma\rightarrow\phi B$, arising 
from the chiral connection, $B\gamma\rightarrow B$,
$\phi\gamma\rightarrow\phi$ or via the Kroll-Rudermann interaction $\gamma B 
\to B \phi$, stemming from the chiral vielbein. These interactions give rise 
to nine different topologies presented in Fig.~\ref{fig:topo}.
Following the conventions of \cite{Berends:1967vi} the most general Lorentz covariant form of the amplitude for the process: $\gamma (k) B_i(p-k)\longrightarrow B_f(p-q) \phi_f(q)$ can be written as $T_{fi}=i\epsilon_\mu\bar u_f\big(\sum_{i=1}^8\mathcal{B}_i\mathcal{N}_i^\mu\big) u_i$, where $\epsilon_\mu$ is the photon polarization vector and $u_{i,f}$ are the initital and final dirac spinors, respectively, which are normalized like $\bar u u=2m$. Moreover $\mathcal{B}$ denote the coefficients of the pseudo-vectors $\mathcal{N}^\mu~\epsilon~\{\gamma_5\gamma^\mu\slashed k,2\gamma_5 P^\mu,2\gamma_5q^\mu,2\gamma_5k^\mu,\gamma_5\gamma^\mu,\gamma_5\slashed k P^\mu ,\gamma_5\slashed k k^\mu,\gamma_5\slashed k q^\mu \}$ with $P=2p-q-k$.

For further study we fix the axis of quantization to the $z$-axis and rewrite the amplitude once more in terms of CGLN amplitudes and Pauli spinors and martices as
	\begin{align}
	T_{fi}=8\pi\sqrt{s}~\chi_f^\dagger\mathcal{F}\chi_i,
	\end{align}
where $\mathcal{F}=i(\sigma\cdot \epsilon)\mathcal{F}_1 + (\sigma\cdot\hat q)(\sigma\cdot[\hat k\times\epsilon])~\mathcal{F}_2 +i(\sigma\cdot\hat k)(\hat q\cdot \epsilon)\mathcal{F}_3+ i(\sigma\cdot\hat q)(\hat q\cdot \epsilon)\mathcal{F}_4 +i(\sigma\cdot \hat k )(\hat k \cdot \epsilon) \mathcal{F}_5+i(\sigma\cdot\hat q) (\hat k\cdot \epsilon) \mathcal{F}_6 +i(\sigma\cdot \hat q)(\hat k\cdot \epsilon)\mathcal{F}_6-i(\sigma\cdot\hat q)\epsilon_0\mathcal{F}_7-i(\sigma\cdot\hat k)\epsilon_0\mathcal{F}_8$ with $\hat k$ and $\hat q$ normalized three-vectors. For the exact form of CGLN amplitudes in form of coefficients $\mathcal{B}$ we refer the reader to the \cite{Borasoy:2007ku}. Let us note here that due to current conservation two of the eight CGLN amplitudes can be expressed in terms of other six. Moreover two of the remaining six CGLN amplitudes are accompanied by scalar components of $\epsilon$ only and thus have no influence on process including real photons. In view on photoproduction this leaves us with four independent CGLN amplitudes.

Let us also mention  that by construction of the unitary hadronic interaction
the photoproduction amplitude obeys the requirement of two-body unitarity in 
the subspace of meson-baryon channels automatically. There are five different 
unitarity classes, which obey the two-body unitarity by themselves. Gauge 
invariance is, however, fulfilled only if all topologies are taken into account.
Note that within the approximation used here, crossing symmetry is violated.

\medskip

\noindent{\textbf{Results}:}~
\begin{figure*}[thb]
\includegraphics[trim=20mm 0mm 20mm 0mm, angle=-90,width=1
\linewidth]{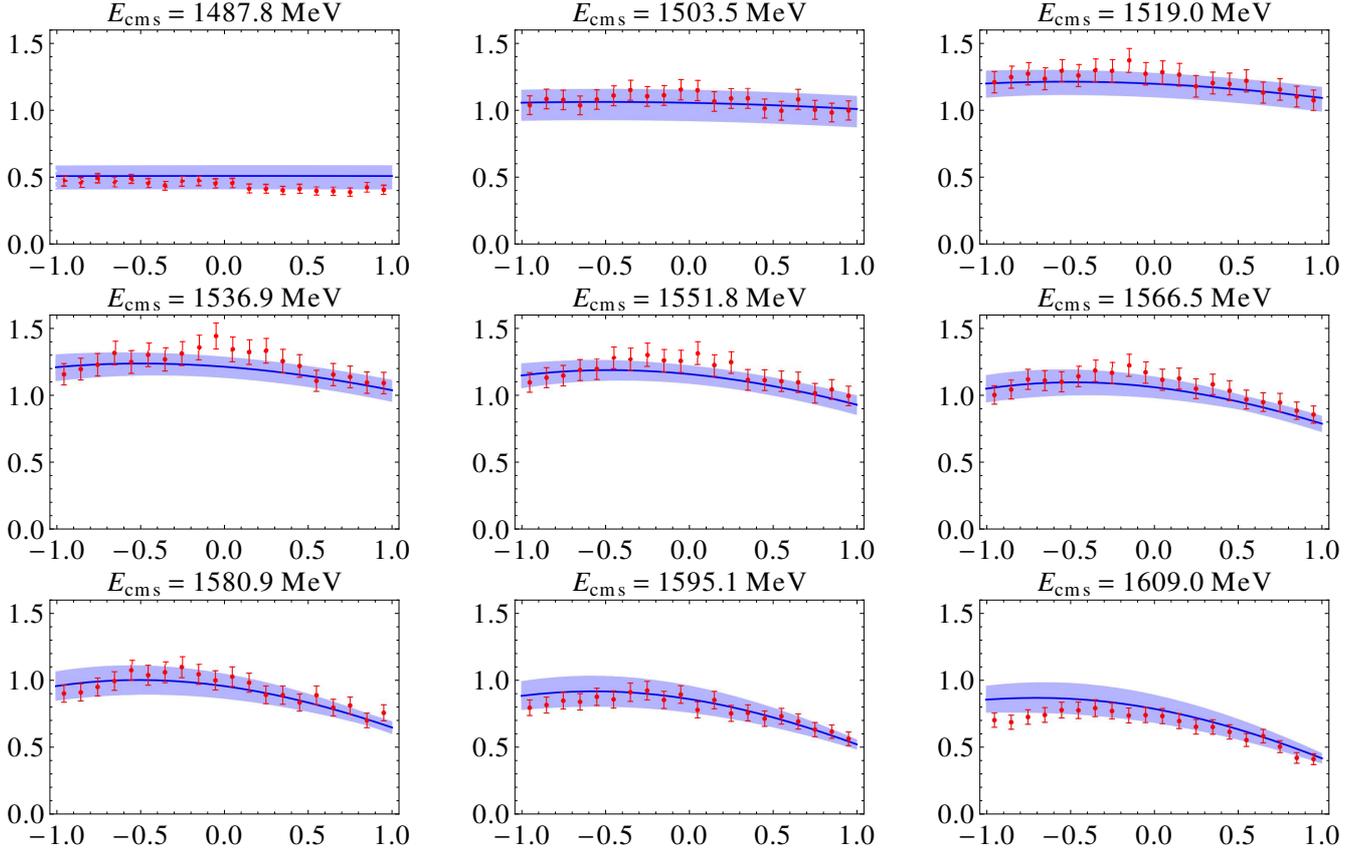}
\caption{Best fit of our model (solid lines with shaded error bands) 
compared to the experimental data from the Crystal Ball at MAMI 
(filled circles with error bars). The differential cross section  
${d\sigma}/{d\Omega}$ in $[\mu b/sr]$ is plotted versus $\cos(\theta)$ 
for the cms energies given on top of each panel.}\label{fig:dsigma}
\end{figure*}
We are now able to confront our model with the experimental results, 
for which we consider the recent measurement of the differential cross 
sections from the Crystal Ball at MAMI~\cite{McNicoll:2010qk}. The parameters of 
the model are the unknown renormalization scales, whose number is restricted 
to four due to isospin-symmetry: $\{\mu_{\pi N}$, $\mu_{\eta N}$, $\mu_{K
  \Lambda}$, $\mu_{K\Sigma}\}$. In contrast to \cite{Borasoy:2007ku} we do 
not consider the meson decay constants as free parameters, but fix them to 
their physical values. Our input parameters are:
 $F_\pi=F_\eta/1.3=0.0924$, $~F_K =0.113$, 
 $M_{\pi^0}=0.135$, $M_{\pi^+}=0.1396$,	
 $M_\eta=0.5478$, $M_{K^+}=0.4937$,
 $M_{K^0}=0.4977$, $m_p=0.9383$,
 $m_n=0.9396$, $m_\Lambda=1.1157$,
 $m_{\Sigma^0}=1.1926$, $m_{\Sigma^+}=1.1894$, $D=0.8$ and $F=0.46$ (all 
masses and decay constants in units of GeV).

The p- and d-waves become  important already at moderate energies away 
from the threshold. Since the kernel of the BSE considered here is the WT-term 
only, which mainly produces an s-wave contribution, we restrict the
center-of-mass (cms) energy of the present analysis to
$1487.8\text{ MeV}<E_{\rm cms}<1609.9\text{ MeV}$. Note that the WT interaction 
kernel produces a small p-wave contribution via the lower components of the 
Dirac spinors, however the explicit angular dependence of the amplitude is
only described when higher order potentials are taken into 
account \cite{CaroRamon:1999jf}. The
calculation of the BSE with the  potential at next-to-leading order is performed 
in Ref.~\cite{Bruns:2010sv} for the hadronic part and is to be extended to 
photoproduction in Ref.~\cite{Mai:2011xx}. 
There the $\eta$-photoproduction off the neutron, which is beyond the scope of 
this work, will also be investigated.

\begin{figure}[t1]
\includegraphics[trim=20mm 0mm 40mm 0mm, angle=-90,width=1
\linewidth]{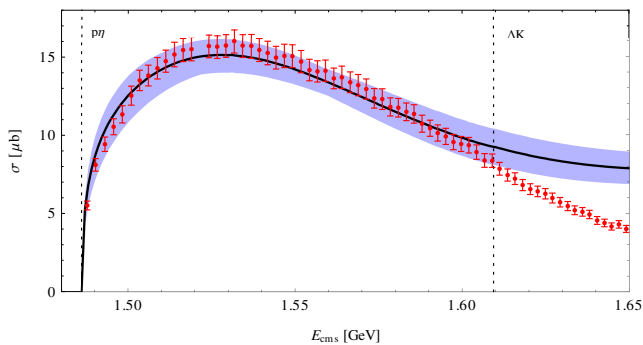}
\caption{Total cross section for the best fit of our model (solid lines with shaded error bands) 
compared to the experimental data from the Crystal Ball at MAMI 
(filled circles with error bars). The vertical dashed lines denote the 
$p\eta$ and the $K\Lambda$ thresholds.}\label{fig:sigma}
\end{figure}

The quality of our  fits is given in terms of the $\chi^2_{\rm dof}$,
	\begin{align}
	\chi^2_{\rm dof}=
	\frac{N}{\sigma N -\delta}
	\sum_{E}\frac{1}{n(E)}
		\sum_z\frac{(\frac{d\sigma}{d\Omega}(E,z;\mu)-\frac{d\sigma}{d\Omega}(E,z;e))^2}
		{(\Delta\frac{d\sigma}{d\Omega}(E,z; e))^2},\nonumber
	\end{align}
where $z=\cos(\theta)$, $n(E)$ is the number of data points at energy $E$,
$\sigma$ is the number of distinct energies and $N$ is the total number of
data points. Moreover $\delta=4$ denotes the number of degrees of freedom, 
which are called collectively $\mu$. The letter $e$ denotes the
experimental values. Using now a random walk minimization procedure,
we obtain a $\chi^2_{\rm dof}=0.9997$ with the following parameters
	\begin{align*}
	\log(\mu_{\pi N}/\text{GeV})&=-0.611,
	&\log(\mu_{\eta N}/\text{GeV})=-0.512^{+0.057}_{-0.051},\\
	\log(\mu_{K \Sigma}/\text{GeV})&=+1.845,
	&\log(\mu_{K \Lambda}/\text{GeV})=-5.112^{+0.403}_{-0.312},
	\end{align*}
where the error bars are obtained by varying the parameters 
such that the $\chi^2_{\rm dof}$ is increased by one. No uncertainty
is given, if it does not affect the value of a parameter at the
given accuracy. Some representative differential cross sections 
are shown in Fig.~\ref{fig:dsigma}. Quite a good agreement
between the model and experiment is achieved for energies more than
$100$~MeV above threshold, which can be seen from the plot of the integrated 
cross section in Fig.~\ref{fig:sigma}. Above the $K\Lambda$ threshold, the
total cross section exceeds the data, this will eventually be overcome
in a more precise NLO calculation.

It is commonly believed that the first nucleon resonance, the $S_{11} (1535)$ 
saturates the cross section close to the $\eta N$ threshold. Moreover, it is 
known that this state can be understood as dynamically generated already from the 
leading chiral order vertex, i.e. the WT-interaction. Since the latter is also the  
driving term of our hadronic amplitude it is worth to have a look at the modulus 
of the electric dipole amplitude $E_{0+}$ as a function of $s=E_{\rm cms}^2$ on the 
second Riemann sheet. The resonance appears at
 	\begin{equation}
 	E_{\rm cms}= \left(1525.9 ^{+4.4}_{-3.6} -  i 111.4^{+1.9}_{-2.0}\right)~\text{MeV}
 	\end{equation}
which is in good agreement with the extraction from phenomenological or other coupled-channel 
approaches 
collected  in \cite{Nakamura:2010zzi} (at least for the real part) 
and our recent determination 
from scattering data \cite{Bruns:2010sv}. It can therefore be identified with the 
$S_{11}(1535)$ resonance, which is dynamically generated in the present approach.
The imaginary part of the pole position is larger than the values of more recent 
phenomenological approaches collected in \cite{Nakamura:2010zzi}, 
however the uncertainty given above reflects only the influence of the errors on 
the parameters and is certainly underestimated. Moreover this large width is consistent
with our determination from scattering at NLO.

For the s-wave multipole $E_{0+}$ we obtain at threshold
 	\begin{equation}
 	E_{0+}~=~\left(-12.39^{+1.51}_{-1.05}+i\text{
          }16.15^{+2.23}_{-1.85}\right)   \cdot 10^{-3}/M_{\pi^+}~.
 	\end{equation}
Its modulus $|E_{0+}| = 20.35^{+2.41}_{-2.38} \cdot 
10^{-3}/M_{\pi^+}~$ is somewhat larger than the one obtained in  the early calculation in unitarized 
chiral perturbation theory  \cite{Kaiser:1996js} and in certain  resonance models 
including the $S_{11}(1535)$, see e.g. Ref.~\cite{Krusche:1995nv}. 
Moreover, the ratio of the imaginary to the real part is about $1.06 \ldots 1.68$ and agrees with 
estimations from  resonance models~\cite{Krusche:1995nv}.

\noindent{\textbf{Extension of the model}:}~
As stated before, one should include all NLO contributions in the hadronic
as well as the electromagnetic part of the production amplitude. 
However, the full inclusion of such terms is beyond the scope of this work 
and will be given later \cite{Bruns:2010sv,Mai:2011xx}. 
As a first step to study  the impact of such a modification, it is interesting
to consider the magnetic moment couplings of the photon to the nucleon, since
these play an important role in phenomenological studies. The corresponding
terms of the NLO chiral Lagrangian read \cite{Kubis:2000aa}
	\begin{align*}
	b_{12}\langle \bar{B}\sigma^{\mu\nu}\{F^+_{\mu\nu},B\}\rangle+
	b_{13}\langle \bar{B}\sigma^{\mu\nu}[F^+_{\mu\nu},B]\rangle ~,
	\end{align*}
where $F^+_{\mu\nu}=uF_{\mu\nu}u^\dagger+u^\dagger F_{\mu\nu}u$ and 
$F_{\mu\nu}=\partial^\mu v^\nu-\partial^\nu v^\mu$ is the
photon field strength tensor. The corresponding additional vertices are 
restricted to $\gamma{}B~\to~B$ and $\gamma{}B{}\phi~\to~B\phi$ only, thus they do 
not give rise to any change of the hadronic part of our photoproduction
amplitude. As a matter of fact unitarity in the subspace of meson-baryon 
channels is still preserved and the gauge invariance is guaranteed
automatically by the functional form of the interaction itself. 

Within our model the inclusion of the $b_{12,13}$ terms leads to an additional 
complication due to the $\Lambda\leftrightarrow\Sigma^0$ transition induced by this
terms. Without going into details, see \cite{Bruns:2010sv,Mai:2011xx} for
discussion, we have to modify the regularization scheme, such that the pure
baryon integrals are set to zero from the beginning. Thus, our treatment of the
loop integrals is, in effect, similar to the EOMS regularization scheme
advocated  in ref.~\cite{Fuchs:2003qc}. Moreover, the same transition implies 
that $\mu_{K}=\mu_{K \Sigma }=\mu_{K \Lambda }$, which reduces the original
parameter  space of our model to $\{\mu_{\pi N}$,~$\mu_{\eta N}$,~$\mu_{K}$,~$b_{12}$,~$b_{13}\}$.

We have performed fits to the Crystal Ball data on differential cross sections 
in the same manner as before. However the energy region in question is reduced 
since no good agreement was achieved even for moderate energies. For the
energy region $1487.8\text{MeV}<E_{\rm cms}<1541.8\text{MeV}$ we can  obtain
a  fit which minimizes the $\chi^2_{\rm dof}$ to $1.75$. For the same
regularization  scheme, but without the inclusion of $b_{12,13}$ terms 
we obtain $\chi^2_{\rm dof}=3.44$. However, the agreement with the data is 
worse compared to the the leading order approach. The conclusion  to be
drawn is that the inclusion of the Pauli-terms $\sim b_{12,13}$  alone 
does not improve the LO description. In fact, it is mandatory to perform
a full NLO analysis. In particular, as the Pauli coupling is much for
important for the neutron, we refrain from giving the LO results for
$\gamma n \to \eta n$ here. 

\medskip

\noindent{\textbf{Summary and outlook}:}~
We have shown that the precise data of $\eta$-photoproduction in the threshold
region can be described rather accurately within the gauge-invariant 
chiral unitary framework. The $S_{11} (1535)$ resonance is
generated dynamically and its pole position in the complex plane agrees with 
earlier determinations. To go to higher energies, to improve
the precision and to investigate also the interesting properties of the 
same reaction on the neutron, one has to go to NLO and, in the long run, also
include the Born terms in the unitarization scheme.  Such efforts are under 
way \cite{Mai:2011xx}.

\medskip

\noindent{\textbf{Acknowledgments}}

We are grateful to Peter Bruns for his stimulating remarks and cooperation. We 
thank Michael D\"oring for useful discussions. 
This work is supported in part by
the DFG (SFB/TR 16 ``Subnuclear Structure of Matter'' and SFB/TR 55 
``Hadron Physics from Lattice QCD''), by the EU HadronPhysics2 project
``Study of Strongly Interacting Matter'', and by the Helmholtz Association 
through funds provided to the Virtual Institute ``Spin and strong QCD'' (VH-VI-231).

\medskip


\end{document}